# DISTRIBUTED RANDOM NUMBER GENERATION FOR THE NEEDS OF PUBLIC GOVERNANCE

Angelin Lalev[1]

*We propose distributed protocol for generation of random numbers via computer systems. The protocol is specifically designed to fit the needs of random selection as it is performed in public sphere and is inspired by real problems, which are posing difficulties for Bulgarian judicial system. Random selection in public sphere is meant to be mechanism for increasing the transparency and reducing the possibilities of collusion between various government actors. Usage of computers reduces transparency and when done improperly, can lead to disastrous consequences concerning the public trust in the institutions of government. This protocol tries to provide technical solution to the transparency problem by allowing third parties to guarantee for the fairness of the random selection without giving these parties the ability to influence maliciously the result.*

## 1. Introduction

Random selection is long been used for breaking ties in tender procedures, choosing members of juries, commissions and panels of experts. The sought after effect of applying random selection is to guarantee and reinforce the fairness of a process that requires a governing body to make impartial decisions. Precisely because the need of transparency, computers so far have been rarely used to aid such random selection by providing random numbers. Yet there are applications where computers cannot be excluded because the volume of selections requires great many random numbers that other methods cannot provide.

Usage of computers to perform random selection always invokes certain questions about the transparency of whole process. In more developed democracies, manipulating computer programs and hardware to influence random selection quite unthinkable and the general public is well disposed to ignore such questions. In less developed democracies though, this mechanism is prone to abuse, and often results exactly the opposite effect on public opinion than one envisioned.

Such was the case with random selection, implemented in Bulgarian judicial system several years ago. From 2007 onwards, the random distribution of court cases to judges is handled by electronic means and is mandated by Bulgarian law. This mechanism was created to address public fears that judges were assigned to some court cases by their superiors based on their personal and political loyalties, which would guarantee certain outcome of the case. It was only at the end of 2014 when software solutions for the implementation of this policy became the focus of public scrutiny. An investigation into the software solution used by majority of Bulgarian courts – LawChoice - concluded that, being standalone application, the solution provided little or no protection from manipulation, even by technically unsophisticated users. It was enough to delete the data files of the program to remove any traces of a choice being made trough the program. This allowed simple strategy for manipulation - the data files would be backed, and then the choice would be made. If it is unfavourable, the data files will be restored and the process will be repeated until favourable result is reached. The efforts to fix the software were directed towards centralization of the random selection and making deletion of the data file without traces harder for the users. The measures were reassuring for the general public, but from technical point of view, the solution remains inadequate because of the fact, that the program and the data still resides on single computer that can be manipulated by the party that has administrative control over it no matter how sophisticated software measures are developed to prevent this.

[1] Dept. of Computer Science, „D.A.Tsenov" Academy of Economics, Svishtov, Bulgaria; lalev@uni-svishtov.bg,

We propose distributed solution to the problem of random selection that closely mimics the techniques used by popular distributed software on Internet. This solution itself is by no means new idea (it has been first investigated by Manuel Blum[2][3] circa 1982), but since then it has been studied and applied more thoroughly only in technical context. We would like to argue that its application in social context has additional aspects that make some of the shortcomings of the general solution irrelevant.

Given the now entrenched public distrust and allegations of corruption at higher levels of the judicial system in Bulgaria (and possibly many other problems, related to corruption in many different countries), this solution is not only mental exercise. It can be implemented on a scale, which would allow it to serve all other random selections just as easily, becoming somewhat similar to what certificate authorities are to electronic signatures today.

The proposed protocol guarantees the generated numbers will be fair even if $n-1$ of all $n$ participants form a coalition and act maliciously. That is, if even one of the participants adheres to the protocol fairly, the rest of participants cannot influence the whole process. Then the fairness of the process in concrete settings will be guaranteed by distributing the duties of generation to parties that are least likely to act in collusion.

The aforementioned property of the protocol is subject to one very important assumption – all participants will be forced to finish the protocol to its end. This assumption is unreasonable when applied to autonomous distributed systems on Internet, where even Byzantine-style errors cannot be excluded, but is more realistic in situations where external arbitrage may be requested and applied, as is the case with social systems.

## 2. Building blocks

The problem of random selection from a list of $k$ elements (as well as ordering such list) can be easily reduced to the generation of uniformly distributed integers in the interval $[0; k)$. To create a scheme that distributes the process of random number generation, one must solve two principal problems:

1) A technique must be found to combine the inputs from each participant. The result has to be a number in the desired interval and has to behave as if drawn from uniform distribution. This must work even if $n-1$ of the total $n$ participants form a coalition and coordinate their inputs with the goal of biasing the results.

2) A technique must be found to ensure that no coalition of $n-1$ of total $n$ participants will be able to exploit the inevitable tiny timing lags in data transmission to introduce bias into the result. For example if all $n-1$ participants in the coalition intentionally wait until the only remaining participant announces their input, this should not allow them to manipulate their own inputs in coordination to influence the resulting number.

**The first problem** is usually solved in practice by making each party generate its own random number in the desired interval. Merging these can then be accomplished by use of binary XOR[4]. XORing multiple random variants, each drawn from a different distribution, ensures that the result is uniformly distributed, given that at least one of these distributions is uniform. So, if even a sole participant draws their random numbers uniformly, the resulting number will also be uniformly distributed. This, of course, is correct only if the desired interval for the random numbers is between 0 and some power of 2, which ensures that XORing will not produce out of range numbers. Producing out of range numbers is not an issue regarding bias if they are dealt with properly (i.e. discarded), but it can introduce unnecessary complexity in the networking communication, which is undesirable for such a protocol. Luckily, this property of XOR follows from a more general result.

As binary operation, XOR is associative and commutative and has the additional properties that $a \oplus a = 0$ and $a \oplus 0 = a$ for any number $a$. So if $a$ and $b$ are numbers with a length of $t$ bits in binary representation, then the relation $\tau_b$, $a \mapsto a \oplus b$ with domain and codomain $S = \{0,1\}^t$ is a function. Then

$$\tau_b(\tau_b(a)) = (a \oplus b) \oplus b = a \oplus (b \oplus b) = a \oplus 0 = a$$

for any $a$, so $\tau_b$ is its own left and right inverse. Since it has a left and right inverse, $\tau_b$ is a bijection from S onto S, hence a permutation. So, if $b$ is drawn uniformly, $\tau_b$ is not only a permutation of $S$, but it is uniform, that is, given the choice of $b$ being uniform, any possible permutation of S is equally likely to occur.

The described property of XOR (namely to make the resulting distribution uniform), follows from the fact that it induces uniformly generated random permutation. If $a$ is an integer in the interval $[0; k)$ and $\phi$ is a (uniformly generated) random permutation of the integers in the same interval, then the probability that $\phi(a)$ equals a particular integer $b \in [0; k)$ does not depend on $a$ and $b$ and is equal to $\frac{1}{k}$. So if $a_1, a_2, \ldots a_p$ are drawn according to arbitrary distribution and $\phi_1, \phi_2, \ldots, \phi_p$ are random permutations, then $\phi_1(a_1), \phi_2(a_2), \ldots, \phi_p(a_p)$ are uniformly distributed.

Thus, if one participant in the protocol generates an initial integer in the interval [0;k), and the other parties generate permutations of the integers in the same interval, then the resulting number $\phi_1(\phi_2(\ldots \phi_{n-1}(a)))$ is uniformly distributed between the integers in $[0; k)$, given that either:
  1) A is randomly drawn from uniform distribution;
  2) One of the permutations $\phi_1, \phi_2 \ldots \phi_n$ is generated uniformly at random;
This property exactly prescribes the operation of the protocol in respect to the generation and combination of each participant's inputs. One participant will start with a uniformly distributed integer in the desired interval and the other participants will provide random uniformly-generated permutations of the integers in the same interval. Unlike XORing, this will always produce an integer in the desired range and will avoid unnecessary bouts of networking communication.

This approach has one great caveat. When each participant in the protocol reveals their input, it has to occur at the exact moment of time as all the other participants reveal theirs. Otherwise, one or more of the participants may cheat the system by deferring the generation and submission of their inputs until they know the inputs of all other participants. They can then generate new inputs that will allow them to totally subvert the choice. This is, in essence, **the second principal issue that the protocol has to overcome**.

Since such simultaneous transmission is absolutely impossible in a networked environment, another solution has to be found. Luckily, the solution is available in the form of commitment schemes. Modern commitment schemes use industry-standard one-way cryptographic hash functions and usually proceed in two phases.

**In the first phase**, each participant generates an input (in our case, a random number) and then uses cryptographic hash function to compute the hash sum of this number. Next, *only* the hash sums of the inputs are exchanged between the parties. When all the participants have all hashes, the commitment scheme goes into its **second phase** and each party reveals the actual original input to the other. Each participant then can check the inputs against the hash sums, verifying that no participant has reneged on its chosen input.

Cryptographic hash functions have three important properties that ensure that the above commitment scheme works correctly and binds each party to its choice, thus preventing dishonesty.

1) Cryptographic hash functions are one-way in the sense that they exhibit *first preimage resistance.* That is, given a hash sum, it is computationally infeasible to find an input (preimage) to the hash function that will result as an output of exactly the given sum. Thus, a participant in our protocol, having only a hash sum from another participant, cannot easily deduce the value that the other participant has committed to.

2) Cryptographic hash functions also exhibit *collision resistance and second preimage resistance*. Second preimage resistance guarantees that, if one has an input to the cryptographic hash function and the corresponding output – the hash sum of this input – it is computationally infeasible to find a *second*, different input that has the same hash sum. *Collision resistance* enhances this property even further, guaranteeing that two values with the same hash sum cannot be efficiently generated together by using some efficient meet-in-the-middle search approach. Thus, if a participant in our protocol generates a number and distributes their hash sum amongst the other participants, it will be impossible for them to generate (at that moment, or later) another number having the same hash sum. Therefore, second preimage resistance and collision resistance ensure that the participant is bound to their generated value once their hash is transmitted to the other participants. This efficiently solves the second major problem, as outlined above.

The protocol has to deal with some additional issues concerning actual implementation. The significance of these problems can be described as less important only due the fact that they are well known, and have widely-available industry-standard implementations.

Technologies for electronic signatures are indispensable to the protocol, assuring the authenticity of the messages. Current industry-standard solutions for their implementation are based on elliptic or Edwards curve cryptography like ECDSA[1] and EdDSA[5].

Random number generators used by the participants **are a highly dangerous problematic source** for the protocol. In case there is a perfect (or true) unbiased and uncorrelated random number generator, the above conclusions regarding the security of the protocol do hold true. In the real world though, random number generators act deterministically, have slight biases, and produce auto-correlated outputs. All these properties might be exploited by an attacker to gain advantage and subvert the protocol. The protocol has the disadvantage that each participant voluntarily reveals much of the output generated by their random number generator to all other participants. This does not immediately constitute a critical weakness, but invites such attacks. To avoid these and achieve a highly unpredictable sequence of bits, the participants should use hardware random number generators. Based on quantum or thermodynamic phenomena, these generators produce truly random output. This output may be XORed with the output of a strong pseudo-random generator such as Blum Blum Shub[3] or a block cipher in counter mode. This will strengthen the protocol even more against such attacks.

### 3. The protocol

Our protocol assumes that the parties have agreed beforehand on a public signature scheme. This assumption is not unreasonable, given that the number of participants will be low enough to solve this problem offline and by direct contact. We also assume that all the participants' public keys are properly exchanged.

The protocol envisions two roles for the participants. A participant may be an initiator or a guarantor. The initiator has a central role in the protocol and only one of the participants may assume the role. All other participants assume the roles of guarantors. The protocol may be described by the following sequence of steps.

**Step 1:** The initiator determines the upper bound *k* for the generated number and gives each draw a consecutive number. These two values are put into a data structure. The initiator signs this structure (see fig. 1) with their private key and sends it to all guarantors.

| k | draw № | attributes of the initiator's electronic signature |
|---|---|---|

Fig. 1. Data structure sent by the initiator in step 1

**Step 2:** Each guarantor, upon receiving the structure, checks the electronic signature of the initiator and the number of the draw (each guarantor should keep a record of previous draws and therefore has to know the number of each next draw). If the electronic signature and the number are correct, each guarantor signs the data structure with his private key, adding the necessary attributes. (See. fig. 2).

| k | draw № | attributes of the initiator's electronic signature | attributes of the guarantor's electronic signature |
|---|---|---|---|

Fig. 2. Data structure returned to the initiator from each guarantor at step 2

Subsequently, the guarantor returns the signed structure to the initiator. If the electronic signature or the numbers are incorrect, the guarantor stops participation in the further execution of the protocol. This can occur with a signed message indicating an error, or silently, which will result in the protocol's abortion. When actually implemented, all systems should be stopped at this moment and audited so that the reason for the fault can be discovered.

**Step 3:** After the initiator obtains all copies of the structure in step 2, the initiator checks the signatures of the guarantors and aggregates all structures into one, signing it again with its own signature. (see fig. 3).

| k | draw № | attributes of the initiator's electronic signature | attributes of guarantor 1's electronic signature |
|---|---|---|---|
| ... | ... | ... | ... |
| k | draw № | attributes of the initiator's electronic signature | attributes of guarantor *n*'s electronic signature |
| **attributes of the initiator's electronic signature (this signature covers the whole structure)** ||||

Fig. 3. Data structure distributed by the initiator at step 3

This structure is sent to all guarantors and the initiator then continues to step 5.

**Step 4:** Upon receiving the structure, each guarantor checks the electronic signatures. Upon incorrect verification, the guarantor stops any further participation in the draw.

**Step 5:** The initiator generates an integer in the range $[0; k)$ and enters it into data structure, which includes a long enough sequence of bits, chosen at random, to play the role of cryptographic "salt". The number of the draw is also added to this structure.

**The structure remains hidden for the moment,** but on conversion to the contents to bits, the guarantor executes the protocol cryptographic hash function (i.e. SHA-3) chosen for the purpose. The resulting hash sum is signed by the initiator and sent to all guarantors (see fig. 4)

| draw № | hash sum | attributes of the initiator's electronic signature |
|---|---|---|

Fig. 4. Data structure distributed by the initiator to all the guarantors at step 5.

**Step 6:** Upon receiving the structure at step 5, every guarantor signs the structure with their own private key and returns it to the initiator.

| draw № | hash sum | attributes of the initiator's | attributes of the guarantor's |
|---|---|---|---|

| | | electronic signature | electronic signature |
|---|---|---|---|

Fig. 5. Data structure returned from each guarantor to the initiator at step 6

**Step 7:** Upon receiving all copies of the structure from step 6, the initiator aggregates them and signs them once again:

| draw № | hash sum | attributes of the initiator's electronic signature | attributes of guarantor 1's electronic signature |
|---|---|---|---|
| ... | ... | ... | ... |
| draw № | hash sum | attributes of the initiator's electronic signature | attributes of guarantor $n$'s electronic signature |
| **attributes of the initiator's electronic signature (this signature covers the whole structure)** | | | |

Fig 6. Data structure distributed from the initiator at step 7

Afterwards, the initiator sends the structure to all guarantors.

**Step 8:** Upon receiving the structure from step 7, every guarantor checks the electronic signatures. If the signatures do not match, the guarantor stops any further participation in the draw.

**Step 9:** Each guarantor generates a random permutation of the integers in the interval [0;k) and stores it in a data structure that includes the number of the draw and a long enough sequence of bits, chosen at random, to serve as cryptographic "salt". **This structure remains hidden for the moment from the rest of the participants.** The guarantor computes the hash sum of the data in the structure. The guarantor signs the sum and the number of the draw and sends them to the initiator.

| draw № | hash sum | attributes of the guarantor's electronic signature |
|---|---|---|

Fig 7. Data structure sent by each of the guarantors on step 9

**Step 10:** Upon receiving all structures from the guarantors, the initiator aggregates and signs them. Then it is sent to the guarantors.

| draw № | hash sum of guarantor 1 | attributes of guarantor 1's electronic signature |
|---|---|---|
| ... | ... | ... |
| draw № | hash sum of guarantor $n$ | attributes of guarantor $n$'s electronic signature |
| **attributes of the initiator's electronic signature (this signature covers the whole structure)** | | |

Fig 8. Data structure disseminated by the initiator to all guarantors at step 10

**Step 11:** Upon receiving the structure from step 10, each guarantor signs it and then returns it to the initiator.

| draw № | hash sum of guarantor 1 | attributes of guarantor 1's electronic signature |
|---|---|---|
| ... | ... | ... |
| draw № | hash sum of guarantor $n$ | attributes of guarantor $n$'s electronic signature |
| **attributes of the initiator's electronic signature (this signature covers the whole structure)** | | |
| **attributes of the guarantor's electronic signature (this signature covers the whole structure, including the attributes of the initiator's signature)** | | |

Fig 9. Data structure returned from the guarantors to the initiator at step 11

When the initiator has received each copy of the structure from step 11, each party has committed to a chosen number or permutation. Moreover, the hash sums so far are correctly distributed amongst all participants. Each participant can verify this, because a correct structure at step 11 would mean that each guarantor has checked and agreed on the accuracy of all sums and signatures. An error in the protocol up to this moment would mean interrupting the draw without the possibility of any coalition between $n-1$ participants in order to learn the details of the remaining participant's number.

**Step 12:** The initiator reveals the contents of the secret structure from step 5 and sends it to all guarantors, signing it with its electronic signature.

| cryptographic salt | draw № | the initiator's random number | attributes of the initiator's electronic signature |
|---|---|---|---|

Fig 10. Data structure disseminated from the initiator to all guarantors at step 12

**Step 13:** Each guarantor reveals the contents of the secret structure from step 9 and sends it to the initiator, signing it with their electronic signature.

| cryptographic salt | draw № | random permutation of the guarantor | attributes of the guarantor's electronic signature |
|---|---|---|---|

Fig 11. Data structure sent from the guarantors to the initiator at step 13

**Step 14:** Upon receiving all copies of the structure from step 9, the initiator aggregates this structure into one, signs it, and sends it to all guarantors.

| cryptographic salt | draw № | random permutation of guarantor 1 | attributes of guarantor 1's electronic signature |
|---|---|---|---|
| cryptographic salt | draw № | random permutation of guarantor 2 | attributes of guarantor 2's electronic signature |
| ... | ... | ... | ... |
| cryptographic salt | draw № | random permutation of guarantor $n$ | attributes of guarantor $n$'s electronic signature |

Fig 12. Data structure sent from the initiator to each guarantor at step 14

**Step 15:** Each site checks the structure from step 14 and, upon correct verification, generates the permutation $\phi = \phi_1 \circ \phi_2 \circ ... \circ \phi_n$. Applying $\phi$ to the initiator's random number produces the end result – an integer in the interval [0; k).

Each participant should then publish the details of this generated process to a publically accessible register, along with the structures exchanged at each point (this would probably be implemented as a set of web pages). The registers of all participants must all agree on the chosen value. To be more useful, the whole process may be preceded by a round where the lists of judges are agreed on by all parties. It may even be transformed into a scheme that takes into account each judge's load. For example, the uniform numbers that are generated must be used as a source for transformation into another distribution.

4. **Threats to the security of the protocol**

There are several main threats to the security of the protocol:

- an attempt by a participant to renege on a value that was committed to at an earlier stage;
- an attempt to reveal the hidden values at steps 5 and 9 by using dictionary attacks, before the corresponding participant has revealed them to the others;
- refusal to participate in the protocol;
- the replay of traffic from previous draws in an attempt to subvert the protocol.

**The issue of replaying old messages** (replay attacks) concerns the possibility that a participant could present messages from a previous draw - which would have a valid electronic signature - as new. To prevent this, all the structures of the protocol have a draw number and each party has to maintain a register of previous draws. The existence of a draw number ensures security in case of replay attacks only if all parties react correctly to it. Upon receiving an incorrect draw number, the party that has received it must discontinue participation and initiate offline audits to discover the cause and resynchronize the register. This behaviour is reasonable, since an occurrence of such errors during normal operation with several data centres / servers should be reasonably low. It is more probable that such an error would occur far earlier to step 11 of the protocol. After that, it becomes a strong indication of dishonesty.

**Dictionary attacks** rely on the fact that one and the same input to the cryptographic hash function always generates the same hash sum. For example, an SHA3 hash of "1" will always be "c89efdaa54c0f20c7adf612882df0950f5a951637e0307cdcb4c672f298b8bc6". Since the integer numbers that will be used by the protocol do not present a large enough search space, it is not hard for an adversary to calculate hashes of all of these numbers and enter these in a table, known as a "dictionary". Subsequently, when a participant in the protocol announces their hash sum, an adversary may try to perform a reverse lookup in this dictionary to find the number. To prevent this, the search space is expanded by adding a pseudo-random sequence of bits of sufficient length. The ultra-safe threshold is around 512 bits long, which would force the attacker to compute (and store) well over $2^{512}$ different inputs, which is totally infeasible.

**The issue of one party reneging on an input they have committed to** has already been covered when discussing the building blocks of the protocol. It must be noted that the cryptographic hash function used in the protocol has to be strong. MD5, SHA1 and SHA-256 have been seriously weakened by developments in cryptanalysis and ever increasing computing power, therefore it appears that most appropriate cryptographic hash function is currently SHA3.

**The most dangerous threat from a practical point of view is the possibility that a participant could sabotage the protocol before or after committing to a value.** After step 11 of the protocol, each party has committed to a value. Simulating a network or system error after this step would be difficult. If a network error occurs and a number of participants drop out, the protocol could be discontinued and finished offline in the presence of arbiters. Each party already has the committed values and it is reasonable to accept that each party should keep logs of its activities. In addition, we can assume that its data centre should be fault tolerant enough to make it extremely unlikely for a participant to lose their own value, thus rendering offline execution impossible.

Sabotage of the protocol before step 11 would not lead to any possibility for bias in the result, but it would make reaching any result impossible. Single errors of this type are not a serious problem, but continuous errors would prevent the protocol from working at all. When such an event occurs, it can rarely be attributed to one particular participant. For example, $n - 1$ participants can enter a coalition to "frame" the sole remaining participant. This is why, in such cases, audits and inspection should be carried out on every data centre. **This limits the maximum count of participants in the protocol.** It is impossible for such a protocol to be executed with participants in the order of tens or hundreds, because audits would become a logistic nightmare. 3 to 5 participants are probably the optimum here, which is more than enough in the case of the judicial system.

This analysis shows the direction in which practical implementations of the protocol to Bulgarian judicial system can be developed. It is impossible for a system based on the protocol to make each single regional court a participant in the system, so a solution based on the protocol would be centralized, but would rely on several guarantors who can be associated with an executive branch (i.e. the Ministry of Justice), the Ombudsman Institution, the Presidency, EU Institutions and so on.

## 5. Conclusion

The proposed protocol demonstrates how a method that can find such successful application in so purely a technical field as peer-to-peer systems can be applied almost without adaptation to outstanding issues in the field of public governance. It must be noted, though, that practical implementation may run into unexpected caveats. Is it true, for example, that constant sabotage of the protocol is impossible? What if such a system only works when there are accredited arbiters and auditors and immediately afterwards starts misbehaving again, generating constant early aborts (obviously, because some participant tries hard to sabotage and discredit it in its entirety). Is it practically possible to find and accuse such a participant in a way that is indisputable for the general public? Will internal strife and potential non-cooperation in the parties itself be enough to overcome technical efforts to keep the system running?

If such a protocol is to be implemented correctly, a decision should be made as to if each data centre should work with its own software or if all data centres will use one and the same. In the latter case, it would become easier for a party that finds an error in the software to subvert all participants. The most serious unknown may be related to the fact that cryptography is unfamiliar to the general public and the conjectured effect of increasing its trust in the random character of selection may not materialize at all.

These questions regarding the protocol can be answered only if the practical implementation of such a system is attempted. For the moment, this looks unlikely, but nonetheless the possibility of constructing such a system must be mentioned and considered in case current electronic distribution systems continue to fail to deter potential transgressions.